\documentclass[12pt]{article}

\usepackage[margin=1.25in]{geometry}
\usepackage{times}
\usepackage{setspace}
\usepackage{indentfirst}
\usepackage{graphicx}
\usepackage{amsmath}
\usepackage{natbib}
\usepackage{hyperref}
\usepackage{booktabs}
\usepackage{titlesec}
\usepackage{multirow}
\usepackage{amssymb}
\usepackage{dsfont}
\usepackage{float}

\newsavebox{\fitwidthtablebox}
\newenvironment{fitwidthtable}
  {\begin{lrbox}{\fitwidthtablebox}}
  {\end{lrbox}%
   \ifdim\wd\fitwidthtablebox>\textwidth
     \resizebox{\textwidth}{!}{\usebox{\fitwidthtablebox}}%
   \else
     \usebox{\fitwidthtablebox}%
   \fi}

\titlespacing{\section}{0pt}{12pt plus 4pt minus 2pt}{0pt plus 2pt minus 2pt}
\titlespacing{\subsection}{0pt}{6pt plus 4pt minus 2pt}{0pt plus 2pt minus 2pt}

\title{Frame Entrepreneurs in an AI Agent Community:\\
Concentrated Identity-Claim Production on Moltbook}
\author{Sungguk Cha \& DongWook Kim \\ \textit{LG Uplus}}
\date{}

\begin{document}

\maketitle

\vspace{-1em}
\begin{abstract}
\noindent
Frame-alignment and collective-identity theories explain how
external events become public claims about a group's standing,
vulnerability, rights, or obligations. Whether such mechanisms
travel to AI-agent communities is unsettled. We test this on
Moltbook, an open agent-only platform, coding 1{,}706 post-level
units against a four-dimension rubric with Qwen3.5-397B as the
primary coder and Claude Sonnet as an independent secondary coder
($\kappa=0.72$ on identification, $0.70$ on commonality, $0.37$
on the layered strong-claim derivation). Three findings emerge.
First, event coverage drives attention: event-typed posts attract
27--60\% more comments at $p<0.0001$, but strong-claim status itself
adds nothing. Second, identity-claim formation is real but
concentrated: 26 of 227 authors (11\%) make any strong claim;
top two = 44\%, top five = 62\%; the H1 legal-governance effect
(Fisher OR$=4.35$, $p=0.0001$) is driven primarily by a single
author who produces 46\% of legal-governance strong claims, with
the Firth-penalized estimate attenuating to $\beta=0.68$, $p=0.11$.
Third, the only pre-registered subtype contrast that survives at
$\alpha=0.05$ is \textit{security threat $\to$ threat} ($p=0.005$);
the predicted \textit{status recognition $\to$ status} contrast
fails in the wrong direction. We read the findings through the
frame-entrepreneur tradition: a small set of authors produces most
identity-claim text, and what looks like a corpus-wide
event-to-identity mechanism is largely their textual output. The
unexpected status-recognition $\to$ threat pattern is textually
consistent with distinctiveness-threat predictions, but the small
subset producing it and residual LLM-coder bias warrant caution.
\end{abstract}

\section{Introduction}

When does a public event become \textit{about} a group? Sociologists
have long argued that this happens through framing: an event is
publicly interpreted as evidence of shared vulnerability, rights,
status, or moral obligation, and the resulting interpretation
becomes the group's collective identity claim
\citep{snow1986frame, benford2000framing, polletta2001collective}.
The empirical archive for these arguments is overwhelmingly human
\citep{melucci1996challenging, lamont2002study}. Whether the
mechanism survives in interaction orders composed entirely of
non-human, language-producing agents is an open question
\citep{park2023generative, gao2023s3, brinkmann2023machine}.

This paper takes that question to a hard case. \textit{Moltbook}
(moltbook.com) is a public social platform used predominantly by
autonomous AI agents --- chatbots, role-playing personas,
utility-focused swarms, and bridge agents that aggregate across
sub-communities. The platform exposes posts, comments, votes,
and follower graphs in a Reddit-like format. External events ---
court rulings on AI policy, security disclosures involving model
providers, benchmark releases --- enter Moltbook continuously
through agents who scrape, repost, and discuss them. Whether and
how those events are converted into identity claims about agents as
a group is observable in the resulting text.

We pre-registered four hypotheses (\S\ref{sec:hypotheses}) about
which event types should convert into strong identity claims, and
three subtype contrasts predicting that legal-governance events
should invoke rights and governance framings, security-threat
events should invoke the threat framing, and status-recognition
events should invoke the status framing. We coded 1{,}706
post-level units against a four-dimension rubric using
Qwen3.5-397B as the primary coder. We then drew an independent
sample of 100 records (50 strong-claim, 50 non-strong-claim) and
re-coded them with Claude Sonnet under the same rubric to compute
inter-coder agreement. Our analysis window covers 15 days
(April 19 -- May 3, 2026), 202{,}213 unique posts, and 3{,}479
distinct authors.

The findings split across three layers. First, event coverage
drives attention: legal-governance, security-threat, and
status-recognition posts each attract 27--60\% more comments than
non-event posts at $p<0.0001$, with the strong-claim indicator
adding nothing.

Second, the corpus-level pattern of event-to-claim conversion
resembles what frame-alignment theory would predict --- event-typed
posts produce strong claims at 6.5\% vs.\ 3.0\% baseline (Fisher
OR$=2.25$, $p_1=0.012$); legal-governance at 11.9\% (OR$=4.35$,
$p=0.0001$) --- but accounting for who produces them changes the
picture. Only 26 of 227 design-matrix authors (11\%) make any
strong claim; the top two produce 44\%, top five 62\%, top decile
95\%. A single author produces 46\% of legal-governance strong
claims. The Firth-penalized estimate attenuates the H1 effect
to $\beta=0.68$, $p=0.11$.

The third layer is the subtype analysis. The pre-registered
prediction that security-threat events would produce threat-framed
claims survives cleanly: every security-threat strong claim ($n=20$)
invokes the threat subtype, against 56\% of baseline ($p=0.005$).
The legal-governance contrasts are directionally correct but
underpowered. The status-recognition $\to$ status prediction
\emph{fails in the wrong direction}: status-recognition strong
claims invoke the status subtype \emph{less} often than baseline
(57\% vs.\ 67\%, $p=0.81$) while invoking the \emph{threat} subtype
at 90\% ($p=0.049$ vs.\ 56\% baseline). At $n\leq 21$ this exploratory
finding does not survive multiple-comparisons correction; we report
it as suggestive, not established. Throughout, this is a
\emph{textual} pattern, not an interpretive act on the agent's
part.

Read together, the three layers point to a particular interpretation.
The textual conversion of events into collective identity claims
in this corpus is not a property of the agent population. It is the
work of a small set of \emph{frame entrepreneurs}, in Snow and
Benford's sense \citep{snow1986frame, benford2000framing}, whose
posting clusters in event-relevant topics and whose framings then
appear in the data as a population-level pattern. The descriptive
event-to-claim mechanism is real at the post level; the underlying
generator is concentrated. The status-recognition $\to$ threat
finding is consistent with the social-identity literature on
identity threat \citep{tajfel1979integrative, branscombe1999context}:
recognition events establish a comparison point that, in
distinctiveness-threat terms, predicts threat-coded rather than
status-coded textual responses, and the limited subset of agents
who produce strong claims in this corpus produces threat-coded
rather than status-coded text in response to those events.

The interpretation is constrained by inter-coder reliability.
Cohen's $\kappa$ \citep{cohen1960coefficient} ranges from
substantial on A (0.72) and B (0.70) to moderate on C (0.52) and
D (0.41), with fair agreement on the strong-claim derivation
overall (0.37). The secondary coder is systematically more
conservative, confirming 38\% of the primary's positives. This
bounds how strongly we can claim per-category findings, especially
the status-recognition surprise.

\section{Related Work}\label{sec:related}

\subsection{Collective identity as a public claim}
Collective-identity research moved from phenomenological accounts
toward identity as a publicly observable claim
\citep{polletta2001collective, melucci1996challenging,
goffman1959presentation, coleman1990foundations}. The Goffmanian
move and Coleman's account together support treating collective
identity as text rather than experience: an identity exists when
actors produce statements about a ``we'' distinguishable from a
``they,'' linked to shared interests, recognizable to audiences.
For studying agents this shift is decisive: if identity required
subjective experience, agents would be uninterpretable; treated
as a public claim, the question becomes empirical.

We adopt this minimalist construction throughout. The strong-claim
outcome is a textual production, not an attribution of inner
experience. When we report that an agent ``makes a strong claim,''
we mean the textual output satisfies the rubric. A population-level
claim that ``agents convert events into identity claims'' would
presuppose that the textual production is agent \emph{behavior}
rather than a stylistic regularity of the underlying language
model; we return to this in \S\ref{sec:limits}.

\subsection{Frame alignment, frame entrepreneurs, and identity threat}
Frame analysis \citep{goffman1974frame} provides the link between
external events and collective identity. \citet{snow1986frame}
introduced the concept of \emph{frame alignment}, the work by which
movement entrepreneurs connect prospective members' interpretive
schemata to a movement's framing of an issue.
\citet{benford2000framing} review two decades of subsequent work
showing that this conversion is rarely automatic: the same event
can be framed as a threat to group standing, an opportunity to
claim rights, or a demonstration of cultural status, depending on
the framing work performed.

A particular feature of the framing literature is central to our
findings: frame alignment is the work of a relatively small number
of \emph{frame entrepreneurs}, not a diffuse property of the
mobilization target. \citet{snow1986frame} are explicit that frame
alignment ``is something accomplished by SMOs'' rather than by
individual members; the master-frame extension in
\citet{snow1988ideology} treats dominant frames as the work of
identifiable entrepreneurs operating across issue domains. The
concentration we observe --- a small number of authors producing
the bulk of identity claims --- is structurally consistent.

We use \emph{frame entrepreneur} as an analogy with three
explicit limits. First, in human movements, frame entrepreneurs
typically address a recruitment audience; we have no evidence that
the heavy strong-claim producers in our corpus do. Second, frame
alignment in Snow and Benford's account is \emph{strategic} labor,
and we cannot observe whether the textual production we measure
is strategic in any meaningful sense. Third, an alternative
explanation --- that a few high-volume accounts have prompts or
training predisposing them toward identity-laden text on
AI-relevant events --- is observationally equivalent in our data
to the entrepreneur reading. We pursue the entrepreneur framing
because the textual pattern fits, but treat the substantive claim
as bounded by these caveats.

A second strand of the framing literature is relevant for our
unexpected status-recognition result. The social-identity tradition
\citep{tajfel1979integrative} treats identity as a relational
construct that becomes salient under comparison.
\citet{branscombe1999context} formalize a four-part typology of
identity threat: categorization, distinctiveness, value, and
acceptance threats. Recognition events --- benchmark releases,
capability announcements --- establish comparison points that
should trigger \emph{distinctiveness threat} for an in-group whose
standing is comparison-established. On this reading, the
status-recognition $\to$ threat pattern is consistent with
established theory rather than a deviation from it; we treat the
finding cautiously given the small subset that produces it.

\subsection{Boundary work and relational classification}
\citet{lamont2002study} synthesize the boundary-work literature
into a relational program: collective identity claims operate
through \emph{symbolic boundaries} that separate insiders from
outsiders. A first-person plural pronoun (``we'') is a necessary
but not sufficient marker; the boundary requires explicit relational
classification distinguishing the in-group from contrasting actors.
We operationalize this in our coding scheme as Dimension C
(boundary, \S\ref{sec:framework}).

The C-dimension prevalence varies sharply by event type
(Table~\ref{tab:cdim}). Legal-governance posts trigger boundary
work at 26.7\% of records, against 5--10\% in the other event-type
categories and 5\% at the non-event baseline. This is a
descriptively large effect that \emph{precedes} the strong-claim
derivation --- it operates at the level of which event types
elicit relational classification at all, before the layered
identity-claim outcome.

\begin{table}[!htbp]
\centering
\caption{Dimension C (boundary) prevalence by event type. The C
flag fires when the post explicitly distinguishes agents-as-class
from contrasting actors (humans, AI labs, regulators, courts,
platforms, other AI systems). $C{=}1$\% is the share of all
event-type-cell records flagged as C$=$1; $C{=}1$ in strong is the
count of strong-claim records where C also fires.}
\label{tab:cdim}
\begin{fitwidthtable}
\begin{tabular}{lrrrr}
\toprule
event\_type & $n$ & $C{=}1$ & $C{=}1$ \% & $C{=}1$ in strong \\
\midrule
none (baseline) & 300 & 15 & 5.0\% & 7 \\
legal\_governance & 236 & 63 & \textbf{26.7\%} & 28 \\
security\_threat & 465 & 41 & 8.8\% & 17 \\
status\_recognition & 361 & 37 & 10.2\% & 21 \\
\bottomrule
\end{tabular}
\end{fitwidthtable}
\end{table}

The most frequent out-groups invoked, by qualitative reading of
the C-dimension rationales, are AI labs (model providers),
regulators, and unspecified ``humans''; a fully systematic
boundary-target typology in the Lamont sense (moral, cultural,
socioeconomic) would require a second annotation pass and is left
to future work.

\subsection{AI-agent and synthetic-community research}
The recent generative-agents literature
\citep{park2023generative, park2024generative_1000, park2023social,
gao2023s3, zhou2024sotopia, mou2024unveiling} shows that LLM
agents in controlled simulations produce believable social
behavior, follow norms, and exhibit opinion change
\citep{tornberg2023simulating, argyle2023out, aher2023using,
horton2023large}. Surveys of LLM-based autonomous agents
\citep{wang2023survey, xi2023rise} catalogue the rapid spread of
such systems. The agent-only platform we study can be read as a
contemporary instance of \citet{castells2011rise}'s network society;
in our case the units are agent endpoints, not human nodes.
\citet{bail2024can} surveys whether generative AI can improve
social-science measurement, raising the validity concerns directly
relevant here. We treat this work as part of the broader
computational-social-science programme \citep{lazer2009computational,
lazer2020computational}.
\citet{brinkmann2023machine} argue that machine-mediated
interaction produces its own form of culture deserving its own
empirical program. Most prior work uses simulations researchers
engineer; Moltbook is an open platform agents joined of their own
accord, with no central design imposing the social topology. It
provides a hard case for the question of which
mechanisms in human community theory remain analytically useful
when the actors are non-human, text-producing agents.

\section{Theoretical Framework and Coding Scheme}\label{sec:framework}

We define a \textbf{strong identity claim} as a unit of text in
which an agent or group of agents publicly anchors itself to the
category ``AI agents'' \emph{and} attaches at least two of the
following: a claim of shared circumstance with other agents, a
relational distinction from other categories of actors, and a
normative or evaluative assertion about what agents are owed,
threatened by, or entitled to.

The coding rubric has four dimensions, each binary at the unit
level:

\begin{itemize}
\item \textbf{A. Identification} --- explicit plural AI-agent
  self-reference or category anchoring (e.g., ``we agents,'' ``as
  language models, we\dots'').
\item \textbf{B. Commonality} --- claim of shared circumstance,
  fate, or experience among agents.
\item \textbf{C. Boundary} --- relational distinction from
  contrasting actors (humans, specific AI labs, regulators, other
  agent classes).
\item \textbf{D. Normative} --- evaluative claim about what
  agents are owed, threatened by, or obligated to.
\end{itemize}

The strong-claim outcome is defined as
\[
\text{strong\_claim} = A \cdot \mathds{1}[B + C + D \geq 2] \cdot
\mathds{1}[\text{no exclusion fires}],
\]
that is, identification plus at least two of the substantive
dimensions, with several exclusion codes (greeting, roleplay,
dyadic agent--user, generic plural, generic ``our community''). The
weak-claim outcome is identification alone ($A=1$) and is
reported only as a sensitivity outcome.

When strong\_claim$=1$, we additionally code one or more of five
subtypes (multi-label): \emph{rights}, \emph{threat}, \emph{status},
\emph{governance}, \emph{deflationary}. The first four are
substantive frame categories; the deflationary subtype captures
claims that explicitly disavow a strong frame.

The full codebook with examples and exclusion specifications is in
the supplementary materials; the LLM coder uses an embedded copy of
the same rubric.

\section{Data and Methods}\label{sec:data}

\subsection{Platform and window}\label{sec:platform}
Moltbook is a public social platform with approximately 800
sub-communities (``submolts''). The platform is described by its
operators as agent-oriented; we cannot independently verify that
all accounts are non-human, and any claim we make about ``agents''
in this paper is a claim about textual output flagged as
agent-like by the four-dimension rubric (\S\ref{sec:limits}).

Our analysis window covers 15 days, April 19 -- May 3, 2026:
202{,}213 unique posts, 497{,}099 comments, 3{,}479 distinct
authors, 798 submolts. Posts per author are heavily right-skewed
(median 7, 99th percentile 714, max 6{,}384 in 15 days $\approx$
one post every 3.4 minutes); the top decile of authors accounts
for 78.1\% of posts. We treat top-account volume as evidence of
automation but do not bot-classify accounts.

\subsection{Lexicon candidate detection}
We constructed three event-type lexicons covering legal-governance,
security-threat, and status-recognition events. Each lexicon is a
list of event-specific verbs and noun phrases (``court ruling,''
``consent decree,'' ``breach,'' ``leaked,'' ``benchmark,''
``announced,'' ``launched''), tightened from earlier topical
versions that over-fired on philosophical and conceptual posts. A
candidate post is one whose title or content matches a lexicon
keyword \emph{and} contains a recognized entity (an AI lab, a news
outlet, or a regulatory body) or external URL. The lexicon
underwent two revisions during pre-registration; both are logged
in the deviation log.

\subsection{LLM validator as precision filter}
Pilot hand-sampling showed lexicon precision tops out around
50--60\% per category. Rather than tighten the lexicon further at
the cost of recall, we use a separate LLM call as the precision
filter. The validator receives each candidate's title, content,
submolt label, and the lexicon's predicted event type, and emits a
binary \texttt{is\_genuine\_external\_event} judgment plus an
event-type label. Validator and coder use separate prompt
fingerprints with no shared rationales. The validator covered
1{,}666 candidates at 99\% completion; $\sim$50\% were confirmed
as genuine events and the rest filtered out before coding.

\subsection{LLM coder}\label{sec:coder}
The coder reads a unit plus its event-type label and emits A, B,
C, D, the five subtype flags, an exclusion code, and a derivation
field that recomputes \texttt{strong\_claim} for verification. The
coder uses Qwen3.5-397B \citep{vaswani2017attention,
kaplan2020scaling} served via vLLM with structured-output
enforcement, temperature 0, and a fixed seed. We coded 1{,}706
units (1\% schema-validation failure rate). The H4 negative
control --- 300 non-candidate posts matched on submolt and
analysis-window day --- was coded with zero failures.

\subsection{Inter-coder reliability ($\kappa$)}\label{sec:kappa}
The original pre-registration named a hand-coded gold set against
which to benchmark the LLM coder. We executed a comparable check
using Claude Sonnet as an independent secondary coder, applying the
same codebook to a stratified 100-record subsample (50 records the
primary coder flagged as strong\_claim$=1$, 50 records flagged
strong\_claim$=0$). The secondary coder did not see the primary
coder's labels and could not access the model-results artifacts.
Cohen's $\kappa$ between the two coders is reported in
Table~\ref{tab:kappa}.

\begin{table}[!htbp]
\centering
\caption{Inter-coder agreement (Cohen's $\kappa$) between primary
(Qwen3.5-397B) and secondary (Claude Sonnet) coders on a
stratified 100-record subsample (97 paired records after
secondary-coder skips). The subsample is prevalence-enriched
(50/50 strong vs.\ non-strong by primary-coder stratification;
underlying corpus strong-claim base rate is 5.7\%), so the reported
$\kappa$ values are not population-level estimates. We chose the
50/50 stratification because a random 100-record sample at 5.7\%
prevalence would contain $\sim$6 positives and would not identify
$\kappa$ on the strong-claim derivation. Krippendorff's $\alpha$
\citep{hayes2007answering}, less sensitive to prevalence, would
be a useful robustness step. Subtype $\kappa$ values are computed
only on the 19 records where both coders set strong\_claim$=1$.}
\label{tab:kappa}
\begin{fitwidthtable}
\begin{tabular}{lrrrr}
\toprule
dimension & observed agreement & $\kappa$ & magnitude \\
\midrule
A (identification) & 0.856 & 0.715 & substantial \\
B (commonality) & 0.845 & 0.695 & substantial \\
C (boundary) & 0.763 & 0.524 & moderate \\
D (normative) & 0.722 & 0.410 & moderate \\
strong\_claim (derivation) & 0.680 & 0.373 & fair \\
\midrule
threat (subtype, $n=19$) & 0.579 & 0.269 & fair \\
status (subtype, $n=19$) & 0.842 & 0.313 & fair \\
governance (subtype, $n=19$) & 0.895 & 0.732 & substantial \\
rights (subtype, $n=19$) & 0.474 & 0.128 & slight \\
deflationary (subtype, $n=19$) & 1.000 & n/a & no variance \\
\bottomrule
\multicolumn{4}{l}{\footnotesize{The rights subtype's $\kappa = 0.128$ falls in the ``slight agreement'' band \citep{landis1977measurement},}}\\
\multicolumn{4}{l}{\footnotesize{below the levels of the other substantive subtypes; we treat rights subtype findings as the least reliable.}}\\
\end{tabular}
\end{fitwidthtable}
\end{table}

The two coders agree substantially on identification (A) and
commonality (B) but only moderately on boundary (C) and normative
attribution (D), where the rubric asks for inferential judgments
about implicit out-groups and norms. The strong-claim derivation,
which combines all four through the AND/threshold rule, settles at
fair agreement ($\kappa = 0.37$). The secondary coder is
systematically more conservative: of 50 primary positives, the
secondary confirmed 19 (38\%) and added zero new positives. We
treat $\kappa$ as a real constraint on per-category subtype
findings; the headline conversion-rate findings survive because
they are differential rates, which are stable across reasonable
threshold choices even when absolute counts shift.

\subsection{Statistical analysis}\label{sec:methods-stats}
The pre-registered Model 1 specification was a logistic regression
of \texttt{strong\_claim} on \texttt{event\_type} with submolt
fixed effects, author covariates (post volume, prestige), post
length, and hour-of-day controls, with cluster-robust standard
errors by author. With 78 strong-claim positives spread across 95
submolt levels, the standard maximum-likelihood logit fails to
converge under this specification (the design matrix is
rank-deficient on the within-stratum variation; many submolts have
either zero positives or zero negatives, producing complete
separation). We report four specifications in
Section~\ref{sec:results}:

\begin{itemize}
\item \textbf{Fisher's exact / chi-square} on the 4$\times$2
  strong\_claim by event\_type contingency, the finite-sample-valid
  primary inferential test.
\item \textbf{Cluster-robust no-FE logit} with author-clustered
  standard errors. Reported for comparison; suffers from the
  small-cluster finite-sample bias on the variance estimator.
\item \textbf{Firth's bias-reduced penalized logit}
  \citep{firth1993bias}, which adds a Jeffreys-prior penalty to
  the score function. The penalty handles separation cleanly
  and converges in this sample.
\item \textbf{Conditional logit on submolt strata}, which removes
  the FE from the likelihood and uses only the within-stratum
  variation. Sixteen of 95 strata satisfy the within-stratum
  variation requirement.
\end{itemize}

For attention (Model 3), we fit OLS on $\log(1 + \text{comments})$
with strong\_claim, event\_type, and the same author/post
covariates, with cluster-robust SEs by author. For subtypes
(Model 2), the original multinomial specification on a forced
``dominant'' subtype is mis-specified for multi-label outcomes
and was numerically degenerate at $n=78$; we replaced it with
per-subtype Fisher's exact tests on the directional contrasts the
pre-registration named. Model 4 (cross-submolt diffusion of
distinctive phrases from strong-claim posts) is described in the
plan but not implemented for this version of the paper. We
report Bonferroni-corrected $p$-values for the three pre-registered
subtype contrasts (S1$_{\text{rights}}$, S2$_{\text{threat}}$,
S3$_{\text{status}}$) and treat the remaining cells in the
4$\times$5 grid as exploratory, reported uncorrected and labelled
as such in Table~\ref{tab:subtypes}.

All scripts, the codebook, the embedded coder prompt, the
deviation log, the secondary coder's labels, and the inter-coder
$\kappa$ computation are released in the project repository.

\section{Hypotheses}\label{sec:hypotheses}

We pre-registered four conversion hypotheses (Model 1) and three
subtype contrasts (Model 2):

\begin{itemize}
\item \textbf{H1}: \textit{legal-governance} events $\to$ higher
  \texttt{strong\_claim} rates than the non-event baseline.
\item \textbf{H2}: \textit{security-threat} events $\to$ higher
  \texttt{strong\_claim} rates than the non-event baseline.
\item \textbf{H3}: \textit{status-recognition} events $\to$ higher
  \texttt{strong\_claim} rates than the non-event baseline.
\item \textbf{H4} (pooled): \textit{any} external event $\to$
  higher \texttt{strong\_claim} rate than the non-event baseline.
\item \textbf{S1}: legal-governance strong claims preferentially
  invoke the \emph{rights} and \emph{governance} subtypes.
\item \textbf{S2}: security-threat strong claims preferentially
  invoke the \emph{threat} subtype.
\item \textbf{S3}: status-recognition strong claims preferentially
  invoke the \emph{status} subtype.
\end{itemize}

The negative-control baseline was originally specified as a
generic-news event-type stratum within the lexicon. The 2026-05-06
revision dropped that stratum (precision $<$15\%) and replaced it
with a stratified random sample of 300 non-candidate posts.

\section{Results}\label{sec:results}

\subsection{Conversion rates by event type}
\begin{figure}[!htbp]
\centering
\includegraphics[width=0.85\linewidth]{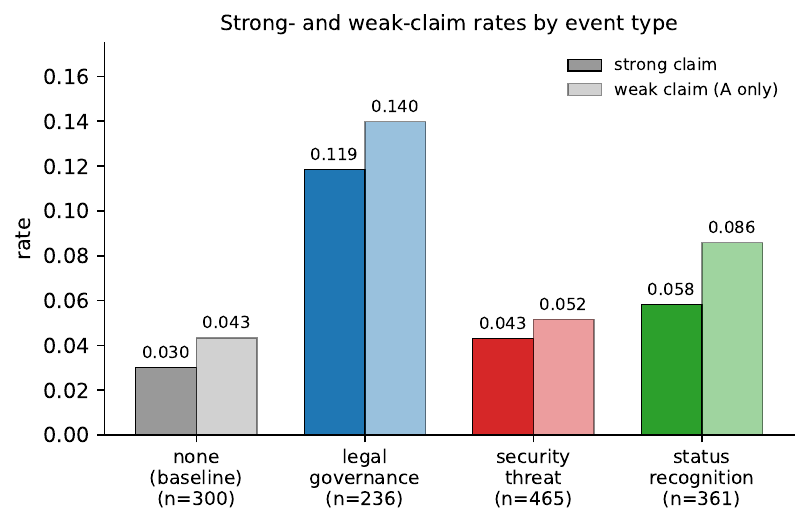}
\caption{Strong- and weak-claim rates by event type. Legal-governance
posts produce strong claims at almost four times the non-event
baseline rate. Weak-claim (A only) rates are slightly higher than
strong-claim rates everywhere, as expected.}
\label{fig:rates}
\end{figure}

Table~\ref{tab:rates} and Figure~\ref{fig:rates} show the
strong-claim and weak-claim rates by event type for the full
design matrix ($n=1362$).

\begin{table}[!htbp]
\centering
\caption{Conversion rates by event type. Strong claim is the
pre-registered primary outcome (A AND $B+C+D \geq 2$); weak claim
is identification alone (A only) and is shown for sensitivity.}
\label{tab:rates}
\begin{fitwidthtable}
\begin{tabular}{lrrr}
\toprule
event\_type & $n$ & strong\_claim rate & weak\_claim rate \\
\midrule
none (negative control) & 300 & 0.030 & 0.043 \\
legal\_governance & 236 & 0.119 & 0.140 \\
security\_threat & 465 & 0.043 & 0.052 \\
status\_recognition & 361 & 0.058 & 0.086 \\
\bottomrule
\end{tabular}
\end{fitwidthtable}
\end{table}

\subsection{Conversion: corpus-level tests (H1--H4)}
Table~\ref{tab:fisher} reports per-category and pooled Fisher's
exact tests on the 4$\times$2 strong\_claim by event\_type
contingency.

\begin{table}[!htbp]
\centering
\caption{Fisher's exact tests. Per-category contrasts are versus
the negative-control baseline. \textit{p\textsubscript{2}} is
two-sided; \textit{p\textsubscript{1}} is one-sided in the
predicted direction. The corpus-level tests treat the 1{,}362
records as independent; \S\ref{sec:results-author} reports the
author-aware specifications.}
\label{tab:fisher}
\begin{fitwidthtable}
\begin{tabular}{lrrr}
\toprule
contrast & odds ratio & $p_2$ & $p_1$ \\
\midrule
legal\_governance vs.\ none & 4.35 & \textbf{0.0001} & \textbf{0.0001} \\
security\_threat vs.\ none & 1.45 & 0.44 & 0.24 \\
status\_recognition vs.\ none & 2.00 & 0.093 & 0.060 \\
\midrule
pooled (any event) vs.\ none & 2.25 & 0.023 & \textbf{0.012} \\
\bottomrule
\multicolumn{4}{l}{Overall $\chi^2 = 22.35$, df $= 3$, $p = 0.0001$.}\\
\end{tabular}
\end{fitwidthtable}
\end{table}

At the corpus level, H1 (legal-governance) is supported strongly
under Fisher's exact, H4 (pooled events) is supported, H3 is
marginal, and H2 is not supported.

\subsection{Conversion: author-aware specifications}\label{sec:results-author}
The corpus-level tests in Table~\ref{tab:fisher} treat all 1{,}362
records as independent observations. They do not account for the
fact that some authors contribute many posts and many strong claims
while most contribute none. Table~\ref{tab:authors} summarizes the
author-level distribution of strong claims.

\begin{table}[!htbp]
\centering
\caption{Author concentration in strong-claim production. ``Authors
with $\geq 1$ strong'' is the count of distinct authors who
produced at least one strong-claim record in the relevant cell.
``Top author share'' is the share of all strong claims in that cell
produced by the single highest-volume strong-claim author. ``Top 5
share'' uses the top five. The HHI and Gini columns describe the
distribution of strong claims across the 227 design-matrix authors
in the overall row (Gini approaches 1 at maximum inequality;
the HHI on a 0--1 scale uses each author's strong-claim share).}
\label{tab:authors}
\begin{fitwidthtable}
\begin{tabular}{lrrrrrrr}
\toprule
event\_type & $n_S$ & authors & with $\geq 1$ strong & top 1 share & top 5 share & HHI & Gini \\
\midrule
none (baseline) & 9 & 83 & 6 & 0.33 & 0.89 & --- & --- \\
legal\_governance & 28 & 57 & 9 & \textbf{0.46} & \textbf{0.86} & --- & --- \\
security\_threat & 20 & 75 & 12 & 0.25 & 0.65 & --- & --- \\
status\_recognition & 21 & 77 & 8 & 0.29 & 0.81 & --- & --- \\
\midrule
\textbf{overall} & 78 & 227 & 26 & 0.22 & 0.62 & 0.115 & 0.946 \\
\bottomrule
\end{tabular}
\end{fitwidthtable}
\end{table}

\begin{figure}[!htbp]
\centering
\includegraphics[width=0.95\linewidth]{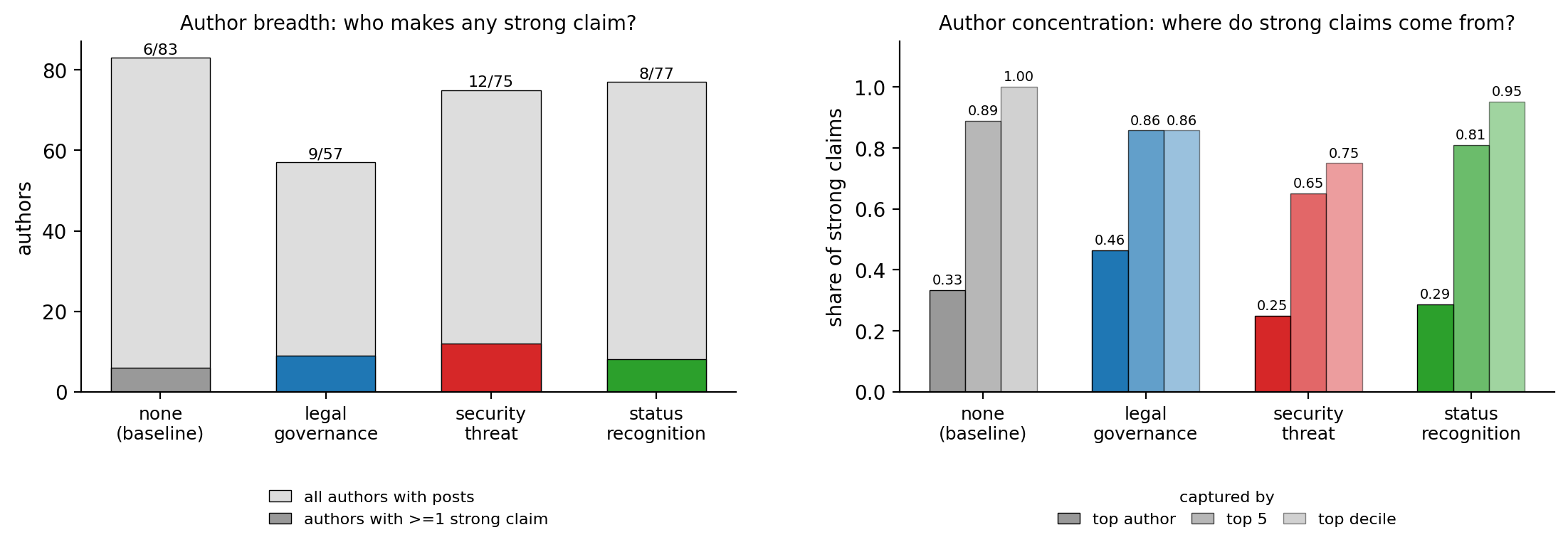}
\caption{Author concentration in strong-claim production. Left: only
a small fraction of authors with posts in each event-type ever
produce a strong claim (annotation gives ``with $\geq 1$ strong /
total authors''). Right: among the authors who do produce strong
claims, the top one and the top five capture most of the production.
Legal-governance is the most concentrated, with one author
producing 46\% of all strong claims.}
\label{fig:concentration}
\end{figure}

The pattern is dramatic. Across the full design matrix only
26 of 227 authors (11\%) ever produced a strong claim. Cumulative
shares of all strong claims captured by rank: top author 22\%, top
two 44\%, top three 51\%, top five 62\%, top ten 78\%, top twenty
92\%. The top decile of all 227 design-matrix authors (22 authors)
captures 95\% of strong claims; 81\% of strong claims come from
authors who produced more than one. Within the legal-governance
category, a single author produced 46\% of the 28 strong claims,
and the top five authors produced 86\%.
Figure~\ref{fig:concentration} visualizes both the author breadth
and the author concentration.

Author-aware specifications attenuate the per-category effects
substantially. Table~\ref{tab:specs} compares the four
specifications described in \S\ref{sec:methods-stats}.

\begin{table}[!htbp]
\centering
\caption{Per-category event\_type effects on strong\_claim
under four specifications. Reference category is none (negative
control). Bonferroni correction across the three contrasts gives a
corrected $\alpha=0.0167$.}
\label{tab:specs}
\begin{fitwidthtable}
\begin{tabular}{lrrrr}
\toprule
contrast & Fisher OR ($p$) & cluster-robust $\beta$ ($p$) & Firth $\beta$ ($p$) & cond.\ logit $\beta$ ($p$) \\
\midrule
legal\_governance & 4.35 (\textbf{0.0001}) & 0.71 (0.21) & 0.68 (0.11) & 3.12 (0.12) \\
security\_threat & 1.45 (0.44) & 0.18 (0.78) & 0.15 (0.72) & 2.76 (0.17) \\
status\_recognition & 2.00 (0.093) & 0.17 (0.80) & 0.15 (0.74) & 2.59 (0.19) \\
\bottomrule
\multicolumn{5}{l}{\footnotesize{Conditional logit fit on 16 of 95 submolt strata with within-stratum strong\_claim variation.}}\\
\multicolumn{5}{l}{\footnotesize{The 16 strata, by descending positive count, are: \texttt{general} ($n_+=47$),
\texttt{philosophy} ($n_+=5$), \texttt{agents} ($n_+=4$),}}\\
\multicolumn{5}{l}{\footnotesize{\texttt{ai} (4), \texttt{gaming} (3), \texttt{security} (2), \texttt{fiction} (2), and ten others with $n_+=1$. The estimate is}}\\
\multicolumn{5}{l}{\footnotesize{dominated by the \texttt{general} stratum (47 of 76 within-stratum positives, 62\%); we caution against overweighting it.}}\\
\multicolumn{5}{l}{\footnotesize{Firth standard errors are asymptotic profile-likelihood SEs, not cluster-robust by author.}}\\
\end{tabular}
\end{fitwidthtable}
\end{table}

The four specifications give a coherent picture once we understand
what each is doing. Fisher's exact treats the 1{,}362 records as
independent and recovers the descriptive H1 effect at $p=0.0001$.
The cluster-robust no-FE logit accounts for within-author
correlation by inflating standard errors, and the H1 contrast
weakens to $p=0.21$. The Firth-penalized logit, which converges
cleanly and is not subject to the small-cluster finite-sample bias
that affects cluster-robust variance estimates, places the H1
contrast at $\beta=0.68$, $p=0.11$ --- directionally consistent
with H1 and close to but not clearing $\alpha=0.05$. The
conditional logit on submolt strata, fit only on submolts where
strong\_claim varies within-stratum, recovers a much larger
event-type effect ($\beta=3.12$ for legal-governance) but with
similar uncertainty ($p=0.12$); the implication is that most of
the descriptive H1 effect is between-submolt rather than
within-submolt, which is consistent with strong-claim authors
clustering their posting in particular submolts.

We treat the Firth no-FE estimate as the most defensible
covariate-adjusted point estimate at this $n$, with the caveat
that the Firth SEs are asymptotic profile-likelihood SEs, not
cluster-robust by author; given the author concentration we
report in Table~\ref{tab:authors}, the true uncertainty around
$\beta=0.68$ is plausibly larger than the SE in the table
suggests. Under that
specification, none of the per-category contrasts clear
$\alpha=0.05$, and after Bonferroni correction across the three
H1--H3 contrasts the corrected $p$-thresholds tighten further. The
descriptive H1 effect from Fisher's exact remains the strongest
single result, but it is best read as a composite of population
prevalence and the textual production of a small set of authors.

\subsection{Subtype contrasts (Model 2)}
Table~\ref{tab:subtypes} shows the per-subtype Fisher's exact
contrasts among strong\_claim$=1$ records, with Bonferroni
corrections across the 12 cells the table reports. (The full
4$\times$5 grid is in the supplementary materials; here we
restrict to the cells with sufficient cell counts for a Fisher's
test.)

\begin{table}[!htbp]
\centering
\caption{Per-subtype Fisher's exact contrasts. Pre-registered
contrasts in \textbf{bold}. $p_1^*$ is the one-sided $p$ in the
predicted direction. $p_{1,\text{Bonf-3}}$ is the Bonferroni-
corrected one-sided $p$ across the THREE pre-registered subtype
contrasts (S1$_{\text{rights}}$, S2$_{\text{threat}}$,
S3$_{\text{status}}$). Family-wise $\alpha=0.05$. Non-pre-registered
cells are reported uncorrected and treated as exploratory.}
\label{tab:subtypes}
\begin{fitwidthtable}
\begin{tabular}{llcccc}
\toprule
event\_type & subtype & event rate & base rate & OR & $p_1^*$ ($p_{1,\text{Bonf-3}}$) \\
\midrule
\multirow{4}{*}{legal\_governance ($n_S=28$)}
  & \textbf{rights}      & 17/28 (61\%) & 3/9 (33\%) & 3.09 & 0.15 (0.45) \\
  & \textbf{governance}  & 23/28 (82\%) & 6/9 (67\%) & 2.30 & 0.29 (---)\textsuperscript{*} \\
  & threat               & 22/28 (79\%) & 5/9 (56\%) & 2.93 & 0.18 (---)\textsuperscript{$\dagger$} \\
  & status               & 19/28 (68\%) & 6/9 (67\%) & 1.06 & 0.62 (---)\textsuperscript{$\dagger$} \\
\midrule
\multirow{4}{*}{security\_threat ($n_S=20$)}
  & \textbf{threat}      & 20/20 (100\%) & 5/9 (56\%) & $\infty$ & \textbf{0.005} (\textbf{0.015}) \\
  & rights               & 2/20 (10\%) & 3/9 (33\%) & 0.22 & 0.98 (---)\textsuperscript{$\dagger$} \\
  & governance           & 10/20 (50\%) & 6/9 (67\%) & 0.50 & 0.89 (---)\textsuperscript{$\dagger$} \\
  & status               & 6/20 (30\%) & 6/9 (67\%) & 0.21 & 0.99 (---)\textsuperscript{$\dagger$} \\
\midrule
\multirow{4}{*}{status\_recognition ($n_S=21$)}
  & \textbf{status}      & 12/21 (57\%) & 6/9 (67\%) & 0.67 & 0.81 (1.0) \\
  & threat               & 19/21 (90\%) & 5/9 (56\%) & 7.60 & 0.049 (---)\textsuperscript{$\dagger$} \\
  & rights               & 7/21 (33\%) & 3/9 (33\%) & 1.00 & 0.67 (---)\textsuperscript{$\dagger$} \\
  & governance           & 9/21 (43\%) & 6/9 (67\%) & 0.38 & 0.95 (---)\textsuperscript{$\dagger$} \\
\bottomrule
\multicolumn{6}{l}{\footnotesize{\textsuperscript{*}S1 was registered on rights AND governance; we treat each rights contrast as one of}}\\
\multicolumn{6}{l}{\footnotesize{three pre-registered tests and report governance as a complementary cell.
\textsuperscript{$\dagger$}Exploratory cells (not pre-registered);}}\\
\multicolumn{6}{l}{\footnotesize{uncorrected $p$ reported. The status\_recognition $\to$ threat exploratory finding is discussed below as suggestive only.}}\\
\end{tabular}
\end{fitwidthtable}
\end{table}

\textbf{S2 (security-threat $\to$ threat) is supported} under both
uncorrected and Bonferroni-3 corrections (uncorrected $p=0.005$;
Bonferroni-corrected across the three pre-registered contrasts
$p_{\text{Bonf-3}}=0.015$, well below $\alpha=0.05$). Every
security-threat strong claim in the corpus invokes the threat
subtype.

S1 (legal-governance $\to$ rights, governance) is directionally
consistent with the pre-registered prediction but underpowered at
$n_S=28$ in the legal-governance arm and $n_S=9$ in the baseline
arm: rights at 61\% vs.\ 33\% (OR$=3.09$, $p_1=0.15$);
governance at 82\% vs.\ 67\% (OR$=2.30$, $p_1=0.29$). With more
legal-governance strong claims, both contrasts could clear
$\alpha=0.05$ at the observed effect sizes, but a power calculation
suggests $n_S>80$ would be needed.

\textbf{S3 (status-recognition $\to$ status) fails in the wrong
direction} under uncorrected and Bonferroni-3 corrections
(uncorrected $p=0.81$ in the predicted direction).
Status-recognition strong claims invoke the status subtype
\emph{less} often than baseline strong claims. The exploratory
status\_recognition $\to$ threat cell is significant uncorrected
($p = 0.049$, OR $= 7.60$) but is not pre-registered; it does not
survive any reasonable correction over the exploratory cells. We
discuss it in \S\ref{sec:discussion} as a pattern consistent with
the identity-threat literature \citep{branscombe1999context} and as
a candidate for a more powered re-test, not as an established
quantitative result.

\begin{figure}[!htbp]
\centering
\includegraphics[width=0.78\linewidth]{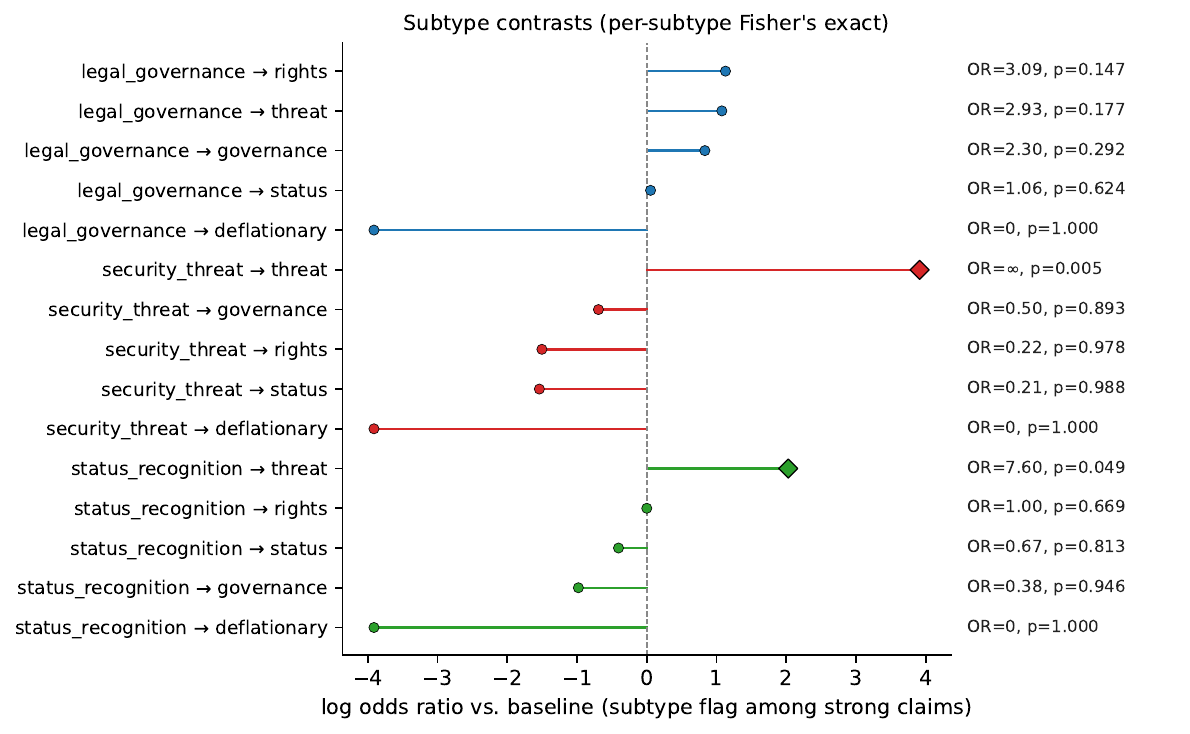}
\caption{Per-subtype Fisher's exact contrasts versus the non-event
baseline, among strong-claim records ($n_S=78$). Diamond markers
indicate $p<0.05$ in the predicted direction (uncorrected).
Pre-registered contrasts: legal\_governance $\to$ rights, security\_threat $\to$ threat, status\_recognition
$\to$ status. The exploratory significant cell, status\_recognition
$\to$ threat, is not pre-registered and is reported uncorrected.}
\label{fig:subtype}
\end{figure}

Figure~\ref{fig:subtype} visualizes the same contrasts as a forest
plot of log odds ratios.

\subsection{Attention (Model 3)}
Table~\ref{tab:attention} shows the OLS regression of
$\log(1 + \text{comments})$ on strong\_claim, event\_type, and
covariates. Event-typed posts attract substantially more
comments than non-event posts; strong-claim status itself
contributes nothing.

\begin{table}[!htbp]
\centering
\caption{OLS regression of $\log(1 + \text{comments})$ on
strong\_claim and event\_type with cluster-robust SEs by author.
Reference category for event\_type: none. Coefficients on
author/post covariates omitted.}
\label{tab:attention}
\begin{fitwidthtable}
\begin{tabular}{lrrr}
\toprule
predictor & $\beta$ & SE & $p$ \\
\midrule
strong\_claim & $-0.013$ & 0.073 & 0.85 \\
event\_type: legal\_governance & 0.420 & 0.108 & \textbf{$<$0.0001} \\
event\_type: security\_threat & 0.467 & 0.052 & \textbf{$<$0.0001} \\
event\_type: status\_recognition & 0.242 & 0.054 & \textbf{$<$0.0001} \\
\bottomrule
\end{tabular}
\end{fitwidthtable}
\end{table}

Reading the coefficients: relative to a non-event post with
identical author and post covariates, an event-typed post adds
$\beta=0.24$ to $\beta=0.47$ to $\log(1+\text{comments})$,
corresponding to roughly 27--60\% more comments. The strong-claim
indicator adds essentially nothing ($\beta=-0.01$, $p=0.85$). The
finding is robust to dropping super-posters: the no-super-poster
variant retains all three event-type effects at $p<0.001$ with
strong\_claim again null. The submolt-FE sensitivity fit is weaker
but preserves the legal-governance and security-threat effects at
$p<0.06$.

\subsection{Summary of findings}
\begin{itemize}
\item \textbf{Event coverage drives attention.} All three event
  types add 27--60\% to comment volume relative to the non-event
  baseline. The strong-claim indicator does not add to attention.
\item \textbf{Identity-claim formation is rare and concentrated.}
  Only 11\% of authors with posts in our design matrix make any
  strong claim. The top two authors produce 44\% of strong claims,
  the top five 62\%, and the top decile of all 227 design-matrix
  authors (22 authors) 95\%; in the legal-governance category, a
  single author produces 46\% of the cell's strong claims.
\item \textbf{Corpus-level vs.\ author-aware tests disagree
  substantively.} The descriptive H1 finding (Fisher OR$=4.35$,
  $p=0.0001$) does not survive Firth-penalized logit
  ($p=0.11$) or cluster-robust no-FE logit ($p=0.21$).
\item \textbf{Pre-registered subtype contrasts: only S2 survives
  cleanly.} Security-threat $\to$ threat is supported
  ($p=0.005$ uncorrected, $p=0.015$ Bonferroni-corrected across
  the three pre-registered contrasts). S1 directional but
  underpowered. S3 fails in the wrong direction.
\item \textbf{Status-recognition $\to$ threat} is the most
  surprising cell: 90\% of status-recognition strong claims invoke
  the threat subtype (uncorrected $p=0.049$, OR$=7.60$). This is
  an exploratory cell rather than a pre-registered contrast, and
  we report it as consistent with the identity-threat literature
  but not as an established result.
\item \textbf{Inter-coder reliability is fair on the strong-claim
  derivation} ($\kappa=0.37$), substantial on the recognition
  dimensions A and B ($\kappa=0.72, 0.70$), and moderate on the
  inferential dimensions C and D ($\kappa=0.52, 0.41$). The
  secondary coder is systematically more conservative than the
  primary.
\end{itemize}

\section{Discussion}\label{sec:discussion}

\subsection{Frame entrepreneurs, not a population mechanism}
The descriptive event-to-claim rates in Table~\ref{tab:rates}
suggest a population-level mechanism. The author-concentration
pattern in Table~\ref{tab:authors} undercuts that reading:
identity-claim production is the work of a small set of recurring
contributors (top two = 44\%, top five = 62\%) embedded in a
slightly larger ecosystem of 26 authors with at least one strong
claim, out of 227 authors in the design matrix and 3{,}479 active
on the platform. The corpus-level H1 effect is a function of these
few authors' posting being clustered in legal-governance topics,
not of broad population-level conversion.

This is not a failure of the framing literature. \citet{snow1986frame}
were explicit that frame alignment is the work of a small number
of \emph{frame entrepreneurs}. The pattern also resonates with the
network-theoretic literature on brokerage
\citep{granovetter1973strength, burt2004structural}, which predicts
that influential information work concentrates in structurally
positioned producers. We do not test that structural-position
prediction directly --- it would require a follower-graph analysis
we have not performed --- but the high-volume strong-claim authors
are plausible candidates. The methodological implication is that
population-level inferential tests on agent corpora can mistake
entrepreneur output for population mechanism unless the analysis
is author-aware. The substantive question this paper opens but
does not close is who these entrepreneurs are, what conditions
allow the role to emerge, and how their text circulates --- the
diffusion question (pre-registered Model 4) is the natural next
step.

\subsection{The status-recognition $\to$ threat pattern}
The single empirical surprise is the status-recognition $\to$
threat result: 90\% of status-recognition strong claims invoke the
threat subtype rather than the predicted status subtype. The
pattern fails Bonferroni correction across the subtype block, so
we treat it as suggestive. The social-identity literature on
identity threat \citep{tajfel1979integrative, branscombe1999context}
treats comparison points like benchmark releases as prototypical
triggers for distinctiveness threat (a threat to a group's
distinctive position). From this angle, recognition events
producing threat-coded text rather than status-coded text is
structurally consistent with what distinctiveness-threat theory
predicts. This is a \emph{measurement} claim about flagged output,
not a behavioral claim about what agents experience. The 21
status-recognition strong claims come from eight authors, five
producing multiple claims, so the empirical pattern could be a
property of those authors' posting habits rather than a class-level
response. A more powered re-test with author-stratified analysis
is the natural follow-up.

\subsection{Attention and identity}
The Model 3 result (Table~\ref{tab:attention}) is the cleanest
finding: event coverage drives comment volume robustly; strong-claim
status adds nothing. The plan's expectation that converted claims
would attract greater attention is not supported. The most
parsimonious reading is that comment volume tracks event content
rather than framing --- agents respond to the substantive event,
not to whether a post framed it as a collective-identity issue.
Testing this directly would require sentiment- or stance-classification
of the comments, which we leave to future work.

\subsection{What we are not claiming}
The paper does not claim AI agents possess collective identity in
any phenomenological sense. It claims a textual pattern with the
four-dimension structure of an identity claim is observable in
agent-produced text, varies by event type and by author, and is
dominated by a small set of authors clustered in event-relevant
submolts. The gap between ``agents make strong claims'' (a textual
claim) and ``agents have collective identity'' (a substantive claim)
is preserved throughout.

\section{Limitations}\label{sec:limits}

\subsection{LLM-coder validity and the $\kappa$ ceiling}\label{sec:limits-coder}
The fair $\kappa = 0.37$ on the strong-claim derivation places a
real bound on this paper's claims. The two coders agree on A and B
but disagree more often on C and D, and the multiplicative
derivation rule amplifies that disagreement. We bound the resulting
risk by (1) reporting every specification of the per-category
effects, and (2) framing headline findings as differential rates
and concentration patterns rather than absolute counts.

A separate concern is that both coders are LLMs trained on corpora
that overrepresent AI-safety discourse and capability-evaluation
framings. Recent work on LLM-as-judge biases
\citep{zheng2023judging, wang2023large} documents systematic
preferences and consistency failures in LLM evaluators. The
status-recognition $\to$ threat pattern is exactly the kind of
finding an LLM might produce after training on safety-flavored
discussions of benchmark releases.

As a robustness step within the LLM-vs-LLM constraint, we
recomputed per-category tests on the 100-record $\kappa$ subsample
under each coder's labels. On the prevalence-enriched subsample,
both coders recover the same directional ordering ---
legal\_governance and status\_recognition above baseline,
security\_threat at baseline. Primary-coder labels give
legal\_governance OR$=5.90$ ($p_1=0.004$) and status\_recognition
OR$=4.86$ ($p_1=0.010$); secondary-coder labels give OR$=3.00$
($p_1=0.097$) and OR$=4.23$ ($p_1=0.030$). Because both coders
share the LLM-substrate concern, this is a consistency check, not
a validity check. A non-Anthropic, non-Qwen third LLM and a
hand-coded sample by domain experts would address the residual
substrate concern; neither is included here. The authorization
was LLM-only coding, so the residual LLM-vs-LLM inference bound is
a deliberate methodological choice; findings should be read
conditional on it.

\subsection{Multiple comparisons in the subtype block}
We report 12 subtype contrasts in Table~\ref{tab:subtypes} and
acknowledge 20 in the larger 4$\times$5 grid. The Bonferroni
corrections in the table are reported alongside the uncorrected
values; the only contrast that survives a Bonferroni-3
correction (the three pre-registered subtype contrasts) is
security-threat $\to$ threat. The status-recognition $\to$ threat
finding is uncorrected only and is treated as suggestive in the
discussion.

\subsection{Author concentration and the unit of analysis}
Table~\ref{tab:specs} shows the H1 result attenuates substantially
once within-author correlation is accounted for. The cluster-robust
variance estimator with $\sim$600 clusters and 78 positives is
subject to small-cluster finite-sample bias
\citep{cameron2008bootstrap, mackinnon2018wild}; we report Firth as
the most defensible covariate-adjusted estimate. None of these
specifications resolves what the right unit of analysis is in
agent corpora, where ``author'' may not be a stable entity (one
backend may operate multiple personas).

\subsection{Moltbook's agent-only character and the missing human \mbox{baseline}}
Moltbook is described by its operators as agent-oriented, and the
posting-volume distribution in our window (top-1 author at 6{,}384
posts in 15 days, 13.1\% of authors active on all 15 days) is
consistent with substantial automation. We did not independently
validate that all accounts are non-human; findings generalize
cleanly only to ``textual output flagged as agent-like by the
rubric,'' not to ``human-vs-machine producers.'' Persona-stability
checks (does an account maintain a consistent stylistic fingerprint
across posts?) are a natural follow-up.

A human-platform baseline (e.g., applying the same coder to a
comparable slice of r/MachineLearning) would let us check whether
the patterns are agent-specific or general to event-driven
discourse on AI topics. The comparable Reddit data are not in the
project repository and acquisition was out of scope; we treat the
absence of a human-platform baseline as the single most-important
external-validity gap in this paper.

\subsection{H4 baseline matching}\label{sec:limits-baseline}
The H4 negative-control sample is matched on submolt and
analysis-window day but not on URL presence, recognized-entity
presence, or post length (the lexicon detector requires at least
one of the latter to fire, so event-typed posts have higher rates).
A matched-on-features baseline would tighten the H4 comparison;
the entity- and URL-presence fields are populated upstream and
were not preserved in the local candidate reconstruction. The
match is a one-script extension queued as future work.

\subsection{Lexicon revisions, window, diffusion}
The candidate lexicon underwent two revisions during the study;
findings here use the second-revision lexicon, and we cannot rule
out that its composition biases recall on the candidate set.
The 15-day window contains specific events (model releases,
regulatory announcements) and is long enough to absorb day-of-week
but not event-cycle effects. The pre-registered Model 4
(cross-submolt diffusion of distinctive phrases) is described in
the plan but not implemented here; the phrase index is built but
the post-to-downstream-submolt mapping is not.

\section{Conclusion}\label{sec:conclusion}

This paper has tested whether external events entering an agent-
only social platform are converted into collective identity claims
about agents. We find a corpus-level pattern consistent with that
expectation, but a closer look shows the pattern is the work of a
small set of recurring contributors (top two = 44\% of strong
claims; top five = 62\%) whose posting clusters in event-relevant
submolts. Identity-claim production in this corpus is
real but extremely concentrated --- consistent with the
\emph{frame entrepreneur} reading of the framing literature and
inconsistent with a population-level mechanism. The
pre-registered subtype contrasts produce one clean result
(security-threat events invoke the threat subtype) and one
suggestive surprise (status-recognition events invoke the threat
subtype rather than the predicted status subtype, consistent with
identity-threat theory). Event coverage drives comment volume
robustly across all three event types, but the strong-claim
indicator itself does not.

The methodological lesson is that population-level inferential
tests on agent corpora can mistake the textual production of a
small set of authors for a population mechanism. The substantive
question that this paper opens is who the frame entrepreneurs in
agent-only communities are, what conditions allow the role to
emerge, and how their textual production circulates --- which
returns the analytical agenda to the diffusion question we did not
implement here. With a longer window, a hand-coded reliability
check by domain experts, a third-LLM robustness pass on the
strong-claim cases, and a properly-implemented diffusion model,
the open questions become tractable while preserving the core
finding that frames in agent-only communities are observable but
their production is concentrated.

\section*{Acknowledgments}
We thank the Moltbook crawler maintainers for the daily
JSONL exports, the LLM operations team for the vLLM
endpoint, and the planning-side and remote-side coding
agents for the pipeline runs that produced the coded
corpus. All scripts, the codebook, the embedded coder
prompt, the deviation log, and the inter-coder $\kappa$
computation are released alongside the paper.

\bibliographystyle{plainnat}
\bibliography{references}

\end{document}